\PassOptionsToPackage{table,dvipsnames}{xcolor}

\documentclass[manuscript,nonacm]{acmart}
\acmConference[EASE 2025]{The 29th International Conference on Evaluation and Assessment in Software Engineering}{17–20 June, 2025}{Istanbul, Türkiye}

\usepackage{mdframed}
\usepackage[shortlabels]{enumitem}
\usepackage{cleveref}
\AtBeginDocument{%
  \providecommand\BibTeX{{%
    \normalfont B\kern-0.5em{\scshape i\kern-0.25em b}\kern-0.8em\TeX}}}
\usepackage{tabularx}

\setcopyright{acmlicensed}
\copyrightyear{2018}
\acmYear{2018}
\acmDOI{XXXXXXX.XXXXXXX}

\acmBooktitle{Woodstock '18: ACM Symposium on Neural Gaze Detection,
 June 03--05, 2018, Woodstock, NY} 
\acmISBN{978-1-4503-XXXX-X/18/06}

\begin{document}

\title{ThreMoLIA: \underline{Thre}at \underline{Mo}deling of \underline{L}arge Language Model-\underline{I}ntegrated \underline{A}pplications}


\author{Felix Viktor Jedrzejewski}
\orcid{0000-0001-7090-2753}
\email{felix.jedrzejewski@bth.se}
\affiliation{%
  \institution{Blekinge Institute of Technology}
  \streetaddress{Valhallavagen 1}
  \city{Karlskrona}
  \state{Blekinge}
  \country{Sweden}
  \postcode{371-41}
}
\author{Davide Fucci}
\orcid{0000-0002-0679-4361}
\email{davide.fucci@bth.se}
\affiliation{%
  \institution{Blekinge Institute of Technology}
  \streetaddress{Valhallavagen 1}
  \city{Karlskrona}
  \state{Blekinge}
  \country{Sweden}
  \postcode{371-41}
}
\author{Oleksandr Adamov}
\orcid{0000-0002-0120-5388}
\email{oleksandr.adamov@bth.se}
\affiliation{%
  \institution{Blekinge Institute of Technology}
  \streetaddress{Valhallavagen 1}
  \city{Karlskrona}
  \state{Blekinge}
  \country{Sweden}
  \postcode{371-41}
}



\begin{abstract}
Large Language Models (LLMs) are currently being integrated into industrial software applications to help users perform more complex tasks in less time.
However, these \textit{L}LM-\textit{I}ntegrated \textit{A}pplications (LIA) expand the attack surface and introduce new kinds of threats.
Threat modeling is commonly used to identify these threats and suggest mitigations.
However, it is a time-consuming practice that requires the involvement of a security practitioner.
Our goals are to 1) provide a method for performing threat modeling for LIAs early in their lifecycle, (2) develop a threat modeling tool that integrates existing threat models, and (3) ensure high-quality threat modeling.
To achieve the goals, we work in collaboration with our industry partner.
Our proposed way of performing threat modeling will benefit industry by requiring fewer security experts' participation and reducing the time spent on this activity. 
Our proposed tool combines LLMs and Retrieval Augmented Generation (RAG) and uses sources such as existing threat models and application architecture repositories to continuously create and update threat models.
We propose to evaluate the tool offline---i.e., using benchmarking---and online with practitioners in the field.
We conducted an early evaluation using ChatGPT on a simple LIA and obtained results that encouraged us to proceed with our research efforts.



\end{abstract}




\maketitle

\section{Introduction}
Large Language Models (LLM) are being integrated into traditional software to enhance their capabilities and performance~\cite{haluza2023artificial}.

However, LLMs extend the threat landscape of an application as attackers can exploit new types of vulnerability~\cite{jiang2023identifying}, such as prompt injections~\cite{greshake2023not}.
Prompt injections can force a model to generate malicious outputs, but carefully crafted instructions---introduced during the post-training phase as guardrails---help mitigate this threat~\cite{dong2024building, welbl2021challenges, gehman2020realtoxicityprompts}.
In the early stages of LLM-Integrated Application (LIA) development, the process of threat modeling facilitates the systematic identification of threats, their risk, as well as corresponding mitigations and their priorities~\cite{myagmar2005threat, turpe2017trouble, konev2022survey}.
Threat modeling also facilitates communicating security risks between stakeholders (e.g., security experts, developers, architects)~\cite{yskout2020threat}. 
A \textit{threat model} is the result of a threat modeling process; it lists the identified threats and corresponding mitigation strategies for a given application~\cite{xiong2019threat}.

However, current threat modeling tools (i.e., Microsoft Threat Modeling Tool, OWASP Threat Dragon, and ThreatModeler) are i) lacking maturity as they require manual efforts and security-related skills~\cite{yskout2020threat} and ii) not adapted to LLM-specific threats.
LIAs contain dynamic and nondeterministic LLM components, which necessitate a new or adapted threat modeling method to identify and mitigate LLM-specific threats.
While the industrial maturity and tool support for threat modeling is low~\cite{yskout2020threat}, the threat models proposed in academia are unsuitable for industrial purposes~\cite{grosse2024towards}.
Practitioners in industries such as telecommunication are recognizing the potential of Artificial Intelligence (AI) and plan to incorporate it in their systems~\cite{britto2023telecom} despite threat modeling methods for LIAs not being validated empirically in industrial contexts.


\begin{mdframed}[backgroundcolor=gray!10,linewidth=2pt, leftline=true, topline=false, bottomline=false, rightline=false]
In this paper, we present our vision of an approach assisting practitioners during the threat modeling of LIAs based on LLM.
\end{mdframed}
Automating threat modeling with an LLM enables security practitioners to accelerate development while maintaining an up-to-date threat model throughout the development lifecycle.
Our approach, developed with an industry partner, prioritizes the quality of threat modeling reports by implementing comprehensive data checks to ensure the accuracy and reliability of the output.
To the best of our knowledge, this paper is the first to report an LLM-supported threat modeling approach for LIAs co-produced with industry.

The rest of this paper is structured as follows.
In Section \ref{sec:BG}, we summarize the research efforts in the area of threat modeling for LLM and describe the components contributing to our adopted threat modeling method for LIAs.
In Section \ref{sec:TTA}, we describe our vision to develop and validate our threat modeling approach in collaboration with our industry partner, including foreseen challenges.
\Cref{sec:CSER} summarizes our preliminary results.
\Cref{sec:conclusion} concludes the paper.

\section{Background}
\label{sec:BG}
This section elaborates on the general concept of threat modeling.

\subsection{Threat Modeling}
The goal of every threat modeling is ``(...) to create an abstraction of the system; profiles of potential attackers, including their goals and methods; and a catalog of potential threats that may arise.''~\cite{shevchenko2018threat}.
Performing threat modeling in an early stage of software development (i.e., architecture and design) provides security practitioners with possibilities to identify and mitigate threats. 
Architects and developers then use these mitigations to improve the software design~\cite{shevchenko2018threat}.
Several threat modeling methods exist whose application is context-dependent.~\cite{shevchenko2018threat}.

According to the well-established practice from Microsoft and OWASP, threat modeling methods use Data-Flow Diagrams (DFDs) to visually represent the system under review by ``(...) offering a high-level yet detailed representation of applications’ architecture(...) and its internal and external data flows''~\cite{schneider2024dataflow}.
There are a variety of DFD styles, but all of them share the same four base item groups---\textit{external entities}, \textit{data flows}, \textit{processes}, and \textit{data stores}~\cite{schneider2024dataflow}.
At the moment, DFDs are the only type of diagram we consider for visualizing the system under review.

STRIDE---i.e., Spoofing, Tampering, Repudiation, Information Disclosure, Denial of Service, Elevation of Privilege---is a prominent threat modeling framework to identify and classify threats in a given software system~\cite{mauri2022modeling}.
It organizes possible threats across five categories---\textit{s}poofing, \textit{t}ampering, \textit{r}epudiation, \textit{i}nformation disclosure, and \textit{d}enial of service.

\subsection{LLM Threat Modeling Frameworks }
While LLMs are rapidly gaining popularity in industry, security researchers and practitioners are discovering new LLM-specific threats~\cite{greshake2023not}.
Frameworks developed by security practitioners, such as OWASP Top 10 for LLM applications\footnote{\url{https://owasp.org/www-project-top-10-for-large-language-model-applications/}} and MITRE ATLAS\footnote{\url{https://atlas.mitre.org/matrices/ATLAS}} are essential to create a threat modeling approach for LIAs. 

In the same fashion as the classic OWASP Top 10, OWASP Top 10 for LLM shows a collection of the most popular security threats encountered in LIAs, emphasizing their consequences, simplicity of exploitation, and common occurrence in real-world applications.
The most frequent LLM threat is Prompt Injection, which represents all attacks where an attacker crafts an input ingested by the target LLM to force the target LLM to misbehave or malfunction~\cite{greshake2023not}.
Based on the high value of training data and its impact on the LLM behavior,  Training Data Poisoning describes inserting manipulated data points into the training data set of a target LLM~\cite{wan2023poisoning}.
Supply Chain Vulnerabilities cover all other threats targeting an LLM through third-party components~\cite{yao2024survey}.
The output of an LLM enables threats, such as Insecure Output (handling), Sensitive Information Disclosure, and Overreliance.
Threats referring more to the model operations are called Model Denial of Service and Excessive Agency.
Model Denial of Service denotes attacks forcing the LLM to increase its latency and energy consumption~\cite{shumailov2021sponge}.
Excessive Agency refers to the threat that an LLM abuses high access privileges in the system it is embedded in~\cite{pankajakshan2024mapping}.
The last threat on the OWASP Top 10 LLM list is called Model Theft. 

MITRE developed and maintains the Adversarial Threat Landscape for Artificial-Intelligence Systems (ATLAS)\footnote{\url{https://atlas.mitre.org/matrices/ATLAS}} to organize threats and mitigations related to generic AI systems.
ATLAS includes 14 high-level tactics (from reconnaissance to exfiltration) used to group 58 specific attack techniques targeting AI components, six of which are specific to LLMs.
\Cref{tab:mapping_of_MITRE_and_OWASP} maps MITRE ATLAS attack techniques to the threats in OWASP's Top 10 for LLM, including the respective high-level tactics.
High-level Tactics describe milestones and rationale for a successful attack\footnote{\url{https://atlas.mitre.org/tactics}}. 
For example, an attacker performs a direct prompt injection to escalate their privileges in the system where the targeted LLM is deployed\footnote{\url{https://atlas.mitre.org/tactics/AML.TA0012}}.
\begin{table*}[]
\rowcolors{2}{gray!15}{white} 
    \begin{tabularx}{\textwidth}{lXX}
   \toprule
    \rowcolor{white} 
       \href{https://atlas.mitre.org/matrices/ATLAS}{MITRE ATLAS} &  \href{https://owasp.org/www-project-top-10-for-large-language-model-applications/}{OWASP Top 10 LLM Ranking} & \href{https://atlas.mitre.org/tactics}{High-Level Tactics} \\
       \toprule
       \href{https://atlas.mitre.org/techniques/AML.T0051}{(direct and indirect) Prompt Injection} & LLM01: Prompt Injection  & \href{https://atlas.mitre.org/tactics/AML.TA0004}{Initial Access}, \href{https://atlas.mitre.org/tactics/AML.TA0006}{Persistence}, \href{https://atlas.mitre.org/tactics/AML.TA0007}{Defense Evasion}, \href{https://atlas.mitre.org/tactics/AML.TA0012}{Privilege Escalation} \\
       
       \href{https://atlas.mitre.org/techniques/AML.T0061}{LLM Prompt Self-Replication}  &  & \href{https://atlas.mitre.org/tactics/AML.TA0006}{Persistence}   \\

       \href{https://atlas.mitre.org/techniques/AML.T0054}{LLM Jailbreak}  &  & \href{https://atlas.mitre.org/tactics/AML.TA0007}{Defense Evasion}, \href{https://atlas.mitre.org/tactics/AML.TA0012}{Privilege Escalation}   \\

       \href{https://atlas.mitre.org/techniques/AML.T0056}{LLM Meta Prompt Extraction}  &  LLM02: Insecure Output Handling &  \href{https://atlas.mitre.org/tactics/AML.TA0008}{Discovery}, \href{https://atlas.mitre.org/tactics/AML.TA0010}{Exfiltration} \\

       \href{https://atlas.mitre.org/techniques/AML.T0062}{Discover LLM Hallucination}  &  &\href{https://atlas.mitre.org/tactics/AML.TA0008}{Discovery}  \\

       \href{https://atlas.mitre.org/techniques/AML.T0062}{LLM Plugin Compromise}  &  LLM07: Insecure Plugin Design &  \href{https://atlas.mitre.org/tactics/AML.TA0012}{Privilege Escalation}, \href{https://atlas.mitre.org/tactics/AML.TA0005}{Execution} \\

        \href{https://atlas.mitre.org/techniques/AML.T0057}{LLM Data Leakage}  
        &  LLM06: Sensitive Information Disclosure &
        \href{https://atlas.mitre.org/tactics/AML.TA0010}{Exfiltration} \\
       \bottomrule
    
    \end{tabularx}
    \caption{Mapping of MITRE Attacks and Tactics to OWASP Top 10 for LLM .}
    \label{tab:mapping_of_MITRE_and_OWASP}
\end{table*} 
We will utilize OWASP Top 10 for LLMs in our threat modeling approach tailored for LIAs and extend the list of threats collected by MITRE ATLAS.


\subsection{Related Work}



Researchers addressed LLM-specific threats using a framework consisting of STRIDE and DREAD (Damage, Reproducibility, Exploitability, Affected Users, and Discoverability) following Shostack’s Four Question Framework~\cite{tete2024threat}.
\citet{derczynski2023assessing} proposes a method for assessing the risk of deploying an LLM with risk cards.
Researchers utilized threat models as guidance for attacks on LIAs in specific contexts.
For example,~\citet{li2024risks} describes how a threat model helped to conduct successful attacks on LLMs integrated into the Smart Grid.
Security researchers successfully attacked a LIA incorporating ChatGPT3.5 and ChatGPT-4o based on the results of an ad-hoc threat model~\cite{jiang2023identifying}.
As shown in~\citet{jiang2023identifying}, a threat model can be operationalized for red-team activities and subsequent mitigation actions---such a use case is formalized in a taxonomy presented in~\cite{verma2024operationalizing}. 
Software engineering security practices applied to LIAs are likely to overlook, mishandle, or ignore Machine Learning (ML) and LLM-specific threats.
\citet{jedrzejewski2024threat} calls for a joint effort between academia and industry.
The authors provide a threat modeling method focusing on common issues security practitioners face in the context of ML systems development.

\section{Our Vision: the ThreMoLIA Approach}
\label{sec:TTA}

\begin{figure*}
    \centering
    \includegraphics[width=1.0\textwidth]{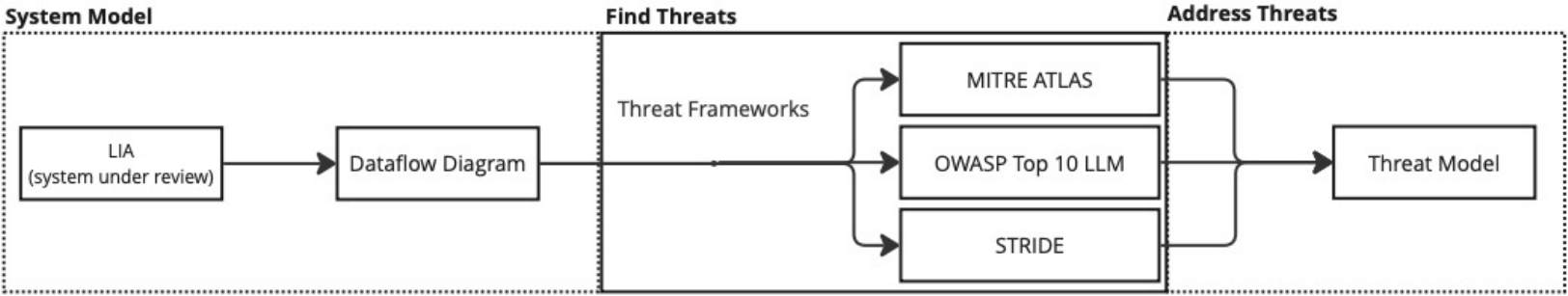}
    \caption{Threat modeling method overview.}
    \label{fig:LIA_Threat_Modeling_Overview}
\end{figure*}

Our vision is to use LLM to support the threat modeling process of LIAs.
Figure \ref{fig:LIA_Threat_Modeling_Overview} shows an overview of the LIA threat modeling approach.
We propose a threat modeling approach guided by~\citet{shostack2014threat}. 
As a first step, the LIA is illustrated by a DFD to create an abstraction of its components and their communication.
This process leverages relevant and available software engineering artifacts, such as the LIA system design documents and requirements specifications provided by the practitioners.
Next, we apply threat frameworks such as MITRE ATLAS, the OWASP Top 10 for LLMs, or STRIDE to identify and assess the most common LLM-specific threats.
Thereafter, the approach requires cataloging each LLM-specific threat impacting the LIA using the attack playbooks and tactics described in MITRE ATLAS.
Furthermore, we plan to include mitigation techniques suggested by MITRE ATLAS to address the identified and assessed threats in the LIA under assessment. 

\subsection{LLM-based Threat Modeling}
\begin{figure*}[h]
    \centering
    \includegraphics[width=0.9\textwidth]{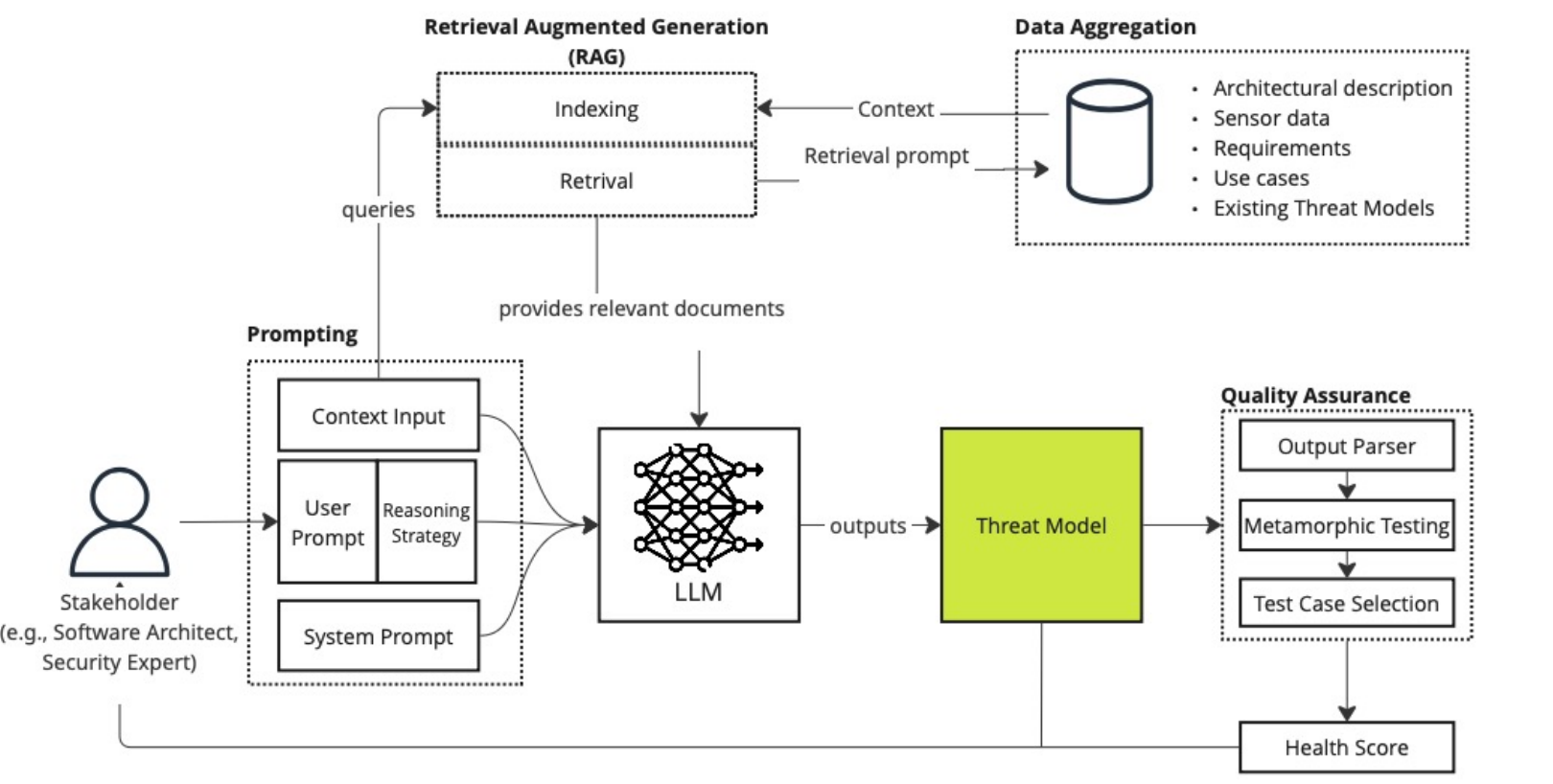}
    \caption{ThreMoLIA overview.}
    \label{fig:ThreMoLIA_Workflow}
\end{figure*}

The overall goal of ThreMoLIA is to generate threat models with the assistance of LLMs, based on stakeholder prompts and relevant data points in the threat modeling process.
Given the non-deterministic character of LLM outputs, the ThreMoLIA approach includes an assessment of the quality of each generated threat model.
We designed ThreMoLIA with composability in mind (see \Cref{fig:ThreMoLIA_Workflow}). 
In this section, we describe its main components and the foreseen challenges in developing and evaluating them. 

\paragraph{Retrieval Augmented Generation (RAG)}
The RAG component provides the LLM with the necessary resources to reason about a system and generate a threat model.
The RAG workflow follows the design summarized by~\citet{gao2023retrieval}, consisting of three stages, indexing, retrieval, and generation.
The incoming query represents the part of the stakeholder request (e.g., a prompt) that provides further context to the LLM.
The RAG component vectorizes the requested information and compares it against the vectorized documents in the database.
Relevant documents are identified based on the closest vectors in the vector space, which represents the vectorized documents from the database.
This component forwards the relevant documents to the LLM.
The Data Aggregation component fills the database with relevant data from different data sources. 

\paragraph{RAG Challenges}
One challenge will be vectorizing graphical document types, as this process is more complex than vectorizing purely text-based documents.
Another point will be to effectively compare documents to discern whether they are relevant to the threat model of a given LIA without introducing redundancy or useless information.

\paragraph{Data Aggregation}
This component collects and aggregates data from different sources, such as natural-language descriptions (e.g., requirement specifications, design documents) and visual representations (e.g., architectural diagrams) of the LIA currently under threat modeling.
Another relevant source is the existing threat models collected from previous threat modeling sessions of the same LIA or from similar projects.
Moreover, once the LIA is in operation, both pre- and post-release, sensor data collected from the monitoring of its components will enable \textit{continuous} threat modeling---e.g., when a new component is added to the architecture, the threat model is automatically regenerated and reviewed in the next stakeholder's threat modeling session.

\paragraph{Data Aggregation Challenges}
The selection of data of suitable quality is an issue.
Stakeholders conducting threat modeling have varying levels of maturity\cite{yskout2020threat}, resulting in heterogeneous data.
Moreover, stakeholders apply company-specific terminology, and there exists a variety of DFD styles~\cite{schneider2024dataflow} which can lead to complications when providing context about the LIA.
Besides data quality, data needs to be prioritized to meet the context constraints of the LLM powering the ThreMoLIA approach. 
If too much data is retrieved, important details may be diluted or lost, and, in the worst case, if the content exceeds the model’s token limit, some parts may be cut off, affecting response accuracy.
Part of the prioritization is to weigh the data sources, as, for example, LLMs are subject to a source bias, preferring content generated by other LLMs\cite{dai2024neural}. 

\paragraph{Prompting}
This component constructs a prompt for the LLM powering ThreMoLIA.
The basic building block is a predefined system prompt engineered to perform the task of threat modeling.
Furthermore, the stakeholders provide a user prompt and a reasoning strategy (e.g., Chain of Thought~\cite{zhang2024k}).
Indications of the context supporting the current system under threat modeling, such as the one provided by the existing DFD and its requirements specification, are also fed to the LLM.
In practice, this context input will trigger the retrieval step in the RAG component.  

\paragraph{Prompting Challenges}
We foresee the key challenge in implementing this component to be the development of a prompt template that allows practitioners with varying degrees of security knowledge to interact with ThreMoLIA reliably---i.e., prompts need to be adapted to different stakeholders' profiles and reflected in the prompting strategy. 
To that end, we will explore conversation disentanglement techniques~\cite{mao2024multi} to handle multiple stakeholder interactions over (possibly) multiple threat modeling sessions. 

\paragraph{Quality Assurance}
This component performs several quality assurance steps on the resulting threat model.
The first step is to check the syntactical correctness of the output that the stakeholder requested in their prompt (e.g., Open Threat Model\footnote{https://github.com/iriusrisk/OpenThreatModel}).

Further, the output is parsed to extract relevant aspects to validate metamorphic relationships~\cite{xie2020mettle}, which represent the oracles used to execute a test suite.
In particular, such a test suite is optimized using different test case selection strategies (e.g., coverage of architectural components).
The quality of the generated threat model is summarized in a health score, further communicated to the stakeholders who can decide to further refine their prompt to improve the quality of the threat model. 

\paragraph{Quality Assurance Challenges}
Currently, established metrics to systematically quantify and evaluate the quality of a threat model are lacking~\cite{yskout2020threat}.
Those metrics are necessary to derive metamorphic relationships specific to the task of threat modeling as well as to define sensible test case selection strategies. 
As a starting point, we consider extending the \textit{Machine Learning Security Maturity Model}\cite{jedrzejewski2023mlsmm}, which proposes a maturity score based on the suggested mitigation techniques applied in a given ML application.

\begin{table*}[t]
\rowcolors{2}{gray!15}{white} 
    \begin{tabularx}{\textwidth}{lX}
   \toprule
   \rowcolor{white} 
       Metric Name & Description \\
    \toprule
            Adversary Capability & Specifies the resources (such as expertise, financial resources, technical resources), methods, and attack vectors. \\
        
        Attack Success Probability (ASP) & Quantifies the likelihood of a successful attack. \\
        
        Exposure Level (EL) & Assesses exposed system vulnerabilities for potential attackers. \\
        
        Impact Severity (IS) & Measures the potential damage or impact of an attack. \\
        
        Likelihood and feasibility of attacks & Calculates the likelihood of attacks besides the feasibility of attacks (non-software-related case). \\
        
        Residual Risk (RR) & Calculates risk after security mitigations have been applied. \\
        
        Identifies phases/elements/patterns in Attacks (composite threats) & Identifies phases, elements, and patterns in attacks, including composite threats. \\
        
        Accuracy &  Represents the number of true positives and true negatives (i.e., correctly identified or omitted threats) produced by the threat modeling tool, divided by the total number of threat classification outcomes. \\
        
        Threat Coverage (TC) & Indicates the proportion of identified threats addressed by the unified threat model. \\
        
        Asset Coverage & Percentage of compromised assets in an attack (simulation). \\
        
        Coverage of MITRE ATT\&CK (tactics and techniques) & Represents a new threat modeling approach (coreLang) mapped against the ATT\&CK matrix as a form of validation, covering 46\%-64\% of ATT\&CK techniques. \\
        
        Threat Library & Describes the knowledge sources used for threat identification. \\

        Scalability & Evaluates the ability to scale the model effectively. \\

        Mitigation Effectiveness & Effectiveness of security controls in reducing identified risks. \\

        Threat Model Reusability & Allows reuse of existing threat models when creating new models. \\
        
        Security Testing & Generates test cases based on the threat model for pentesting. \\
        
        SDLC Integration & Represents the feasibility of applying the threat modeling tool in the software development lifecycle (SDLC) together with other tools. \\

        Model Complexity (MC) & Assesses the complexity of the unified threat model in terms of the number of nodes and relationships. \\

        Visualization of Risk Models & Provides a visual representation of risk for better comprehension. \\
        
        Engineer Friendly & Represents user satisfaction with the threat modeling tool. \\
       
       \bottomrule
    
    \end{tabularx}
    \caption{Preliminary Evaluation Metrics.}
    \label{tab:preliminary_evaluation_metrics}
\end{table*} 
\subsection{Evaluation Plan}
We aim to evaluate ThreMoLIA and assess the efficiency and effectiveness of the stakeholders using it.
There is no agreed-upon benchmark to evaluate a threat model~\cite{yskout2020threat}.
Our first step is to create such a benchmark by extracting proposed and applied metrics through a systematic study of the relevant literature. 
We validate these metrics in focus groups with two security specialists and test their applicability in an industrial case study.

Based on the obtained metrics, we will conduct experiments in industrial settings to compare ThreMoLIA performance with a \textit{traditional} approach---i.e., with the tools currently used by our industrial partner.
We plan to conduct a multiple case study as a final assessment of ThreMoLIA once deployed in an industrial environment. 
The goal is to observe and analyze patterns in multiple products within our industry partner to gain a deeper understanding of how ThreMoLIA fits the workflows of security practitioners.

\section{Current state and early results}
\label{sec:CSER}
This section summarizes the preliminary results we collected.
The initial prototype reflects our first approach to determine whether applying LLMs during threat modeling is promising.
Furthermore, we report a preliminary set of evaluation metrics.

\subsection{Initial Prototype}
We conducted a first investigation of how ChatGPT-3.5 Turbo performs in a zero-shot threat modeling task, using the architectural description of a simple LIA.
ChatGPT referred to the content of OWASP Top 10 LLM and MITRE ATLAS matrix.
In our test, the first prompt contained the request to conduct threat modeling on a system whose architecture we further describe for each component in the same prompt (i.e., without the need to build context from the RAG database).
Since we proposed vague threat modeling instructions, ChatGPT gave us the possibility to either create a threat model for the entire system or for a single component without clarifying which threat modeling method, such as STRIDE, would be applied.
Furthermore, ChatGPT asked to choose between frameworks, such as OWASP Top 10 LLM and MITRE ATLAS, to follow during the threat modeling task.\footnote{\url{https://chat.openai.com/share/7a624c6e-9ec2-4625-9ca3-0f4df8d222cf}}

\subsection{Evaluation Metrics}
To evaluate the ThreMoLIA approach, we investigated the literature to extract applied metrics from other studies evaluating threat modeling approaches.
A focus group consisting of three security researchers evaluated the extracted metrics for their applicability and relevance.
Table \ref{tab:preliminary_evaluation_metrics} shows a preliminary selection of the metrics, we plan to use to evaluate LIA threat models.

In the next step, two security experts will provide feedback based on the selected metrics and potentially add additional ones not covered by the literature.
The metrics will help to evaluate the threat models generated by ThreMoLIA. 

\section{Conclusion}
\label{sec:conclusion}
LLMs can augment the capabilities of software systems and services, but they also introduce a new set of threats that we need to detect and mitigate systematically and reliably.
Researchers and practitioners developed and applied threat modeling methods explicitly tailored to deterministic traditional (i.e., non-AI) software.
LIAs pose a new challenge for existing threat modeling methods, requiring researchers and practitioners to evaluate their effectiveness and either develop new approaches or adjust existing ones. 

In this vision paper, we argue that an approach to support practitioners in threat modeling of LIAs is necessary.  
To that end, we proposed ThreMoLIA, an LLM-based approach focusing on critical aspects such as context input representation and quality assurance for a complex task such as threat modeling.
Moreover, we report the current state of ThreMoLIA development, including an evaluation plan that will be executed in collaboration with an industry partner.

\section{Acknowledgments}
We would like to acknowledge that this work was supported by the KKS foundation through the SERT Research Profile project (research profile grant 2018/010) at Blekinge Institute of Technology and the Threat Modeling for LLM-Integrated applications(ThreMoLIA) Research Project supported by Vinnova (Sweden's Innovation Agency) (Diarienummer 2024-00659).

\bibliographystyle{ACM-Reference-Format}
\bibliography{references}


\begin{thebibliography}{32}


\ifx \showCODEN    \undefined \def \showCODEN     #1{\unskip}     \fi
\ifx \showDOI      \undefined \def \showDOI       #1{#1}\fi
\ifx \showISBNx    \undefined \def \showISBNx     #1{\unskip}     \fi
\ifx \showISBNxiii \undefined \def \showISBNxiii  #1{\unskip}     \fi
\ifx \showISSN     \undefined \def \showISSN      #1{\unskip}     \fi
\ifx \showLCCN     \undefined \def \showLCCN      #1{\unskip}     \fi
\ifx \shownote     \undefined \def \shownote      #1{#1}          \fi
\ifx \showarticletitle \undefined \def \showarticletitle #1{#1}   \fi
\ifx \showURL      \undefined \def \showURL       {\relax}        \fi
\providecommand\bibfield[2]{#2}
\providecommand\bibinfo[2]{#2}
\providecommand\natexlab[1]{#1}
\providecommand\showeprint[2][]{arXiv:#2}

\bibitem[Britto et~al\mbox{.}(2023)]%
        {britto2023telecom}
\bibfield{author}{\bibinfo{person}{Ricardo Britto}, \bibinfo{person}{Timothy Murphy}, \bibinfo{person}{Massimo Iovene}, \bibinfo{person}{Leif Jonsson}, \bibinfo{person}{Melike Erol-Kantarci}, {and} \bibinfo{person}{Benedek Kov{\'a}cs}.} \bibinfo{year}{2023}\natexlab{}.
\newblock \showarticletitle{Telecom AI Native Systems in the Age of Generative AI--An Engineering Perspective}.
\newblock \bibinfo{journal}{\emph{arXiv preprint arXiv:2310.11770}} (\bibinfo{year}{2023}).
\newblock


\bibitem[Dai et~al\mbox{.}(2024)]%
        {dai2024neural}
\bibfield{author}{\bibinfo{person}{Sunhao Dai}, \bibinfo{person}{Yuqi Zhou}, \bibinfo{person}{Liang Pang}, \bibinfo{person}{Weihao Liu}, \bibinfo{person}{Xiaolin Hu}, \bibinfo{person}{Yong Liu}, \bibinfo{person}{Xiao Zhang}, \bibinfo{person}{Gang Wang}, {and} \bibinfo{person}{Jun Xu}.} \bibinfo{year}{2024}\natexlab{}.
\newblock \showarticletitle{Neural retrievers are biased towards llm-generated content}. In \bibinfo{booktitle}{\emph{Proceedings of the 30th ACM SIGKDD Conference on Knowledge Discovery and Data Mining}}. \bibinfo{pages}{526--537}.
\newblock


\bibitem[Derczynski et~al\mbox{.}(2023)]%
        {derczynski2023assessing}
\bibfield{author}{\bibinfo{person}{Leon Derczynski}, \bibinfo{person}{Hannah~Rose Kirk}, \bibinfo{person}{Vidhisha Balachandran}, \bibinfo{person}{Sachin Kumar}, \bibinfo{person}{Yulia Tsvetkov}, \bibinfo{person}{Mark~R Leiser}, {and} \bibinfo{person}{Saif Mohammad}.} \bibinfo{year}{2023}\natexlab{}.
\newblock \showarticletitle{Assessing language model deployment with risk cards}.
\newblock \bibinfo{journal}{\emph{arXiv preprint arXiv:2303.18190}} (\bibinfo{year}{2023}).
\newblock


\bibitem[Dong et~al\mbox{.}(2024)]%
        {dong2024building}
\bibfield{author}{\bibinfo{person}{Yi Dong}, \bibinfo{person}{Ronghui Mu}, \bibinfo{person}{Gaojie Jin}, \bibinfo{person}{Yi Qi}, \bibinfo{person}{Jinwei Hu}, \bibinfo{person}{Xingyu Zhao}, \bibinfo{person}{Jie Meng}, \bibinfo{person}{Wenjie Ruan}, {and} \bibinfo{person}{Xiaowei Huang}.} \bibinfo{year}{2024}\natexlab{}.
\newblock \showarticletitle{Building guardrails for large language models}.
\newblock \bibinfo{journal}{\emph{arXiv preprint arXiv:2402.01822}} (\bibinfo{year}{2024}).
\newblock


\bibitem[Gao et~al\mbox{.}(2023)]%
        {gao2023retrieval}
\bibfield{author}{\bibinfo{person}{Yunfan Gao}, \bibinfo{person}{Yun Xiong}, \bibinfo{person}{Xinyu Gao}, \bibinfo{person}{Kangxiang Jia}, \bibinfo{person}{Jinliu Pan}, \bibinfo{person}{Yuxi Bi}, \bibinfo{person}{Yi Dai}, \bibinfo{person}{Jiawei Sun}, \bibinfo{person}{Haofen Wang}, {and} \bibinfo{person}{Haofen Wang}.} \bibinfo{year}{2023}\natexlab{}.
\newblock \showarticletitle{Retrieval-augmented generation for large language models: A survey}.
\newblock \bibinfo{journal}{\emph{arXiv preprint arXiv:2312.10997}}  \bibinfo{volume}{2} (\bibinfo{year}{2023}).
\newblock


\bibitem[Gehman et~al\mbox{.}(2020)]%
        {gehman2020realtoxicityprompts}
\bibfield{author}{\bibinfo{person}{Samuel Gehman}, \bibinfo{person}{Suchin Gururangan}, \bibinfo{person}{Maarten Sap}, \bibinfo{person}{Yejin Choi}, {and} \bibinfo{person}{Noah~A Smith}.} \bibinfo{year}{2020}\natexlab{}.
\newblock \showarticletitle{Realtoxicityprompts: Evaluating neural toxic degeneration in language models}.
\newblock \bibinfo{journal}{\emph{arXiv preprint arXiv:2009.11462}} (\bibinfo{year}{2020}).
\newblock


\bibitem[Greshake et~al\mbox{.}(2023)]%
        {greshake2023not}
\bibfield{author}{\bibinfo{person}{Kai Greshake}, \bibinfo{person}{Sahar Abdelnabi}, \bibinfo{person}{Shailesh Mishra}, \bibinfo{person}{Christoph Endres}, \bibinfo{person}{Thorsten Holz}, {and} \bibinfo{person}{Mario Fritz}.} \bibinfo{year}{2023}\natexlab{}.
\newblock \showarticletitle{Not what you've signed up for: Compromising real-world llm-integrated applications with indirect prompt injection}. In \bibinfo{booktitle}{\emph{Proceedings of the 16th ACM Workshop on Artificial Intelligence and Security}}. \bibinfo{pages}{79--90}.
\newblock


\bibitem[Grosse et~al\mbox{.}(2024)]%
        {grosse2024towards}
\bibfield{author}{\bibinfo{person}{Kathrin Grosse}, \bibinfo{person}{Lukas Bieringer}, \bibinfo{person}{Tarek~R Besold}, {and} \bibinfo{person}{Alexandre~M Alahi}.} \bibinfo{year}{2024}\natexlab{}.
\newblock \showarticletitle{Towards more Practical Threat Models in Artificial Intelligence Security}. In \bibinfo{booktitle}{\emph{33rd USENIX Security Symposium (USENIX Security 24)}}. \bibinfo{pages}{4891--4908}.
\newblock


\bibitem[Haluza and Jungwirth(2023)]%
        {haluza2023artificial}
\bibfield{author}{\bibinfo{person}{Daniela Haluza} {and} \bibinfo{person}{David Jungwirth}.} \bibinfo{year}{2023}\natexlab{}.
\newblock \showarticletitle{Artificial intelligence and ten societal megatrends: an exploratory study using GPT-3}.
\newblock \bibinfo{journal}{\emph{Systems}} \bibinfo{volume}{11}, \bibinfo{number}{3} (\bibinfo{year}{2023}), \bibinfo{pages}{120}.
\newblock


\bibitem[Jedrzejewski et~al\mbox{.}(2023)]%
        {jedrzejewski2023mlsmm}
\bibfield{author}{\bibinfo{person}{Felix Jedrzejewski}, \bibinfo{person}{Davide Fucci}, {and} \bibinfo{person}{Oleksandr Adamov}.} \bibinfo{year}{2023}\natexlab{}.
\newblock \showarticletitle{MLSMM: Machine Learning Security Maturity Model}.
\newblock \bibinfo{journal}{\emph{arXiv preprint arXiv:2306.16127}} (\bibinfo{year}{2023}).
\newblock


\bibitem[Jedrzejewski(2024)]%
        {jedrzejewski2024threat}
\bibfield{author}{\bibinfo{person}{Felix~Viktor Jedrzejewski}.} \bibinfo{year}{2024}\natexlab{}.
\newblock \showarticletitle{Threat Modeling of ML-intensive Systems: Research Proposal}. In \bibinfo{booktitle}{\emph{Proceedings of the IEEE/ACM 3rd International Conference on AI Engineering-Software Engineering for AI}}. \bibinfo{pages}{264--266}.
\newblock


\bibitem[Jiang et~al\mbox{.}(2023)]%
        {jiang2023identifying}
\bibfield{author}{\bibinfo{person}{Fengqing Jiang}, \bibinfo{person}{Zhangchen Xu}, \bibinfo{person}{Luyao Niu}, \bibinfo{person}{Boxin Wang}, \bibinfo{person}{Jinyuan Jia}, \bibinfo{person}{Bo Li}, {and} \bibinfo{person}{Radha Poovendran}.} \bibinfo{year}{2023}\natexlab{}.
\newblock \showarticletitle{Identifying and mitigating vulnerabilities in llm-integrated applications}.
\newblock \bibinfo{journal}{\emph{arXiv preprint arXiv:2311.16153}} (\bibinfo{year}{2023}).
\newblock


\bibitem[Konev et~al\mbox{.}(2022)]%
        {konev2022survey}
\bibfield{author}{\bibinfo{person}{Anton Konev}, \bibinfo{person}{Alexander Shelupanov}, \bibinfo{person}{Mikhail Kataev}, \bibinfo{person}{Valeriya Ageeva}, {and} \bibinfo{person}{Alina Nabieva}.} \bibinfo{year}{2022}\natexlab{}.
\newblock \showarticletitle{A survey on threat-modeling techniques: protected objects and classification of threats}.
\newblock \bibinfo{journal}{\emph{Symmetry}} \bibinfo{volume}{14}, \bibinfo{number}{3} (\bibinfo{year}{2022}), \bibinfo{pages}{549}.
\newblock


\bibitem[Li et~al\mbox{.}(2024)]%
        {li2024risks}
\bibfield{author}{\bibinfo{person}{Jiangnan Li}, \bibinfo{person}{Yingyuan Yang}, {and} \bibinfo{person}{Jinyuan Sun}.} \bibinfo{year}{2024}\natexlab{}.
\newblock \showarticletitle{Risks of Practicing Large Language Models in Smart Grid: Threat Modeling and Validation}.
\newblock \bibinfo{journal}{\emph{arXiv preprint arXiv:2405.06237}} (\bibinfo{year}{2024}).
\newblock


\bibitem[Mao et~al\mbox{.}(2024)]%
        {mao2024multi}
\bibfield{author}{\bibinfo{person}{Manqing Mao}, \bibinfo{person}{Paishun Ting}, \bibinfo{person}{Yijian Xiang}, \bibinfo{person}{Mingyang Xu}, \bibinfo{person}{Julia Chen}, {and} \bibinfo{person}{Jianzhe Lin}.} \bibinfo{year}{2024}\natexlab{}.
\newblock \showarticletitle{Multi-user chat assistant (MUCA): a framework using LLMS to facilitate group conversations}.
\newblock \bibinfo{journal}{\emph{arXiv preprint arXiv:2401.04883}} (\bibinfo{year}{2024}).
\newblock


\bibitem[Mauri and Damiani(2022)]%
        {mauri2022modeling}
\bibfield{author}{\bibinfo{person}{Lara Mauri} {and} \bibinfo{person}{Ernesto Damiani}.} \bibinfo{year}{2022}\natexlab{}.
\newblock \showarticletitle{Modeling threats to AI-ML systems using STRIDE}.
\newblock \bibinfo{journal}{\emph{Sensors}} \bibinfo{volume}{22}, \bibinfo{number}{17} (\bibinfo{year}{2022}), \bibinfo{pages}{6662}.
\newblock


\bibitem[Myagmar et~al\mbox{.}(2005)]%
        {myagmar2005threat}
\bibfield{author}{\bibinfo{person}{Suvda Myagmar}, \bibinfo{person}{Adam~J Lee}, {and} \bibinfo{person}{William Yurcik}.} \bibinfo{year}{2005}\natexlab{}.
\newblock \showarticletitle{Threat modeling as a basis for security requirements}.
\newblock  (\bibinfo{year}{2005}).
\newblock


\bibitem[Pankajakshan et~al\mbox{.}(2024)]%
        {pankajakshan2024mapping}
\bibfield{author}{\bibinfo{person}{Rahul Pankajakshan}, \bibinfo{person}{Sumitra Biswal}, \bibinfo{person}{Yuvaraj Govindarajulu}, {and} \bibinfo{person}{Gilad Gressel}.} \bibinfo{year}{2024}\natexlab{}.
\newblock \showarticletitle{Mapping LLM Security Landscapes: A Comprehensive Stakeholder Risk Assessment Proposal}.
\newblock \bibinfo{journal}{\emph{arXiv preprint arXiv:2403.13309}} (\bibinfo{year}{2024}).
\newblock


\bibitem[Schneider et~al\mbox{.}(2024)]%
        {schneider2024dataflow}
\bibfield{author}{\bibinfo{person}{Simon Schneider}, \bibinfo{person}{Nicolas E~Diaz Ferreyra}, \bibinfo{person}{Pierre-Jean Queval}, \bibinfo{person}{Georg Simhandl}, \bibinfo{person}{Uwe Zdun}, {and} \bibinfo{person}{Riccardo Scandariato}.} \bibinfo{year}{2024}\natexlab{}.
\newblock \showarticletitle{How Dataflow Diagrams Impact Software Security Analysis: an Empirical Experiment}.
\newblock \bibinfo{journal}{\emph{arXiv preprint arXiv:2401.04446}} (\bibinfo{year}{2024}).
\newblock


\bibitem[Shevchenko et~al\mbox{.}(2018)]%
        {shevchenko2018threat}
\bibfield{author}{\bibinfo{person}{Nataliya Shevchenko}, \bibinfo{person}{Timothy~A Chick}, \bibinfo{person}{Paige O’Riordan}, \bibinfo{person}{Thomas~Patrick Scanlon}, {and} \bibinfo{person}{Carol Woody}.} \bibinfo{year}{2018}\natexlab{}.
\newblock \showarticletitle{Threat modeling: a summary of available methods}.
\newblock \bibinfo{journal}{\emph{Software Engineering Institute| Carnegie Mellon University}} (\bibinfo{year}{2018}).
\newblock


\bibitem[Shostack(2014)]%
        {shostack2014threat}
\bibfield{author}{\bibinfo{person}{Adam Shostack}.} \bibinfo{year}{2014}\natexlab{}.
\newblock \bibinfo{booktitle}{\emph{Threat modeling: Designing for security}}.
\newblock \bibinfo{publisher}{John Wiley \& Sons}.
\newblock


\bibitem[Shumailov et~al\mbox{.}(2021)]%
        {shumailov2021sponge}
\bibfield{author}{\bibinfo{person}{Ilia Shumailov}, \bibinfo{person}{Yiren Zhao}, \bibinfo{person}{Daniel Bates}, \bibinfo{person}{Nicolas Papernot}, \bibinfo{person}{Robert Mullins}, {and} \bibinfo{person}{Ross Anderson}.} \bibinfo{year}{2021}\natexlab{}.
\newblock \showarticletitle{Sponge examples: Energy-latency attacks on neural networks}. In \bibinfo{booktitle}{\emph{2021 IEEE European symposium on security and privacy (EuroS\&P)}}. IEEE, \bibinfo{pages}{212--231}.
\newblock


\bibitem[Tete(2024)]%
        {tete2024threat}
\bibfield{author}{\bibinfo{person}{Stephen~Burabari Tete}.} \bibinfo{year}{2024}\natexlab{}.
\newblock \showarticletitle{Threat Modelling and Risk Analysis for Large Language Model (LLM)-Powered Applications}.
\newblock \bibinfo{journal}{\emph{arXiv preprint arXiv:2406.11007}} (\bibinfo{year}{2024}).
\newblock


\bibitem[T{\"u}rpe(2017)]%
        {turpe2017trouble}
\bibfield{author}{\bibinfo{person}{Sven T{\"u}rpe}.} \bibinfo{year}{2017}\natexlab{}.
\newblock \showarticletitle{The trouble with security requirements}. In \bibinfo{booktitle}{\emph{2017 IEEE 25th International Requirements Engineering Conference (RE)}}. IEEE, \bibinfo{pages}{122--133}.
\newblock


\bibitem[Verma et~al\mbox{.}(2024)]%
        {verma2024operationalizing}
\bibfield{author}{\bibinfo{person}{Apurv Verma}, \bibinfo{person}{Satyapriya Krishna}, \bibinfo{person}{Sebastian Gehrmann}, \bibinfo{person}{Madhavan Seshadri}, \bibinfo{person}{Anu Pradhan}, \bibinfo{person}{Tom Ault}, \bibinfo{person}{Leslie Barrett}, \bibinfo{person}{David Rabinowitz}, \bibinfo{person}{John Doucette}, {and} \bibinfo{person}{NhatHai Phan}.} \bibinfo{year}{2024}\natexlab{}.
\newblock \showarticletitle{Operationalizing a threat model for red-teaming large language models (llms)}.
\newblock \bibinfo{journal}{\emph{arXiv preprint arXiv:2407.14937}} (\bibinfo{year}{2024}).
\newblock


\bibitem[Wan et~al\mbox{.}(2023)]%
        {wan2023poisoning}
\bibfield{author}{\bibinfo{person}{Alexander Wan}, \bibinfo{person}{Eric Wallace}, \bibinfo{person}{Sheng Shen}, {and} \bibinfo{person}{Dan Klein}.} \bibinfo{year}{2023}\natexlab{}.
\newblock \showarticletitle{Poisoning language models during instruction tuning}. In \bibinfo{booktitle}{\emph{International Conference on Machine Learning}}. PMLR, \bibinfo{pages}{35413--35425}.
\newblock


\bibitem[Welbl et~al\mbox{.}(2021)]%
        {welbl2021challenges}
\bibfield{author}{\bibinfo{person}{Johannes Welbl}, \bibinfo{person}{Amelia Glaese}, \bibinfo{person}{Jonathan Uesato}, \bibinfo{person}{Sumanth Dathathri}, \bibinfo{person}{John Mellor}, \bibinfo{person}{Lisa~Anne Hendricks}, \bibinfo{person}{Kirsty Anderson}, \bibinfo{person}{Pushmeet Kohli}, \bibinfo{person}{Ben Coppin}, {and} \bibinfo{person}{Po-Sen Huang}.} \bibinfo{year}{2021}\natexlab{}.
\newblock \showarticletitle{Challenges in detoxifying language models}.
\newblock \bibinfo{journal}{\emph{arXiv preprint arXiv:2109.07445}} (\bibinfo{year}{2021}).
\newblock


\bibitem[Xie et~al\mbox{.}(2020)]%
        {xie2020mettle}
\bibfield{author}{\bibinfo{person}{Xiaoyuan Xie}, \bibinfo{person}{Zhiyi Zhang}, \bibinfo{person}{Tsong~Yueh Chen}, \bibinfo{person}{Yang Liu}, \bibinfo{person}{Pak-Lok Poon}, {and} \bibinfo{person}{Baowen Xu}.} \bibinfo{year}{2020}\natexlab{}.
\newblock \showarticletitle{METTLE: A metamorphic testing approach to assessing and validating unsupervised machine learning systems}.
\newblock \bibinfo{journal}{\emph{IEEE Transactions on Reliability}} \bibinfo{volume}{69}, \bibinfo{number}{4} (\bibinfo{year}{2020}), \bibinfo{pages}{1293--1322}.
\newblock


\bibitem[Xiong and Lagerstr{\"o}m(2019)]%
        {xiong2019threat}
\bibfield{author}{\bibinfo{person}{Wenjun Xiong} {and} \bibinfo{person}{Robert Lagerstr{\"o}m}.} \bibinfo{year}{2019}\natexlab{}.
\newblock \showarticletitle{Threat modeling--A systematic literature review}.
\newblock \bibinfo{journal}{\emph{Computers \& security}}  \bibinfo{volume}{84} (\bibinfo{year}{2019}), \bibinfo{pages}{53--69}.
\newblock


\bibitem[Yao et~al\mbox{.}(2024)]%
        {yao2024survey}
\bibfield{author}{\bibinfo{person}{Yifan Yao}, \bibinfo{person}{Jinhao Duan}, \bibinfo{person}{Kaidi Xu}, \bibinfo{person}{Yuanfang Cai}, \bibinfo{person}{Zhibo Sun}, {and} \bibinfo{person}{Yue Zhang}.} \bibinfo{year}{2024}\natexlab{}.
\newblock \showarticletitle{A survey on large language model (llm) security and privacy: The good, the bad, and the ugly}.
\newblock \bibinfo{journal}{\emph{High-Confidence Computing}} (\bibinfo{year}{2024}), \bibinfo{pages}{100211}.
\newblock


\bibitem[Yskout et~al\mbox{.}(2020)]%
        {yskout2020threat}
\bibfield{author}{\bibinfo{person}{Koen Yskout}, \bibinfo{person}{Thomas Heyman}, \bibinfo{person}{Dimitri Van~Landuyt}, \bibinfo{person}{Laurens Sion}, \bibinfo{person}{Kim Wuyts}, {and} \bibinfo{person}{Wouter Joosen}.} \bibinfo{year}{2020}\natexlab{}.
\newblock \showarticletitle{Threat modeling: from infancy to maturity}. In \bibinfo{booktitle}{\emph{Proceedings of the ACM/IEEE 42nd International Conference on Software Engineering: New Ideas and Emerging Results}}. \bibinfo{pages}{9--12}.
\newblock


\bibitem[Zhang et~al\mbox{.}(2024)]%
        {zhang2024k}
\bibfield{author}{\bibinfo{person}{Yadong Zhang}, \bibinfo{person}{Shaoguang Mao}, \bibinfo{person}{Tao Ge}, \bibinfo{person}{Xun Wang}, \bibinfo{person}{Yan Xia}, \bibinfo{person}{Man Lan}, {and} \bibinfo{person}{Furu Wei}.} \bibinfo{year}{2024}\natexlab{}.
\newblock \showarticletitle{K-Level Reasoning with Large Language Models}.
\newblock \bibinfo{journal}{\emph{arXiv preprint arXiv:2402.01521}} (\bibinfo{year}{2024}).
\newblock


\end{thebibliography}




\end{document}